\documentclass[conference]{IEEEtran}

\ifCLASSINFOpdf

\else

\fi

\usepackage[dvipsnames]{xcolor}
\usepackage{ wasysym }
\usepackage{hyperref}
\usepackage{enumitem}
\newlist{RQ}{enumerate}{1}
\setlist[RQ]{label=\textbf{RQ\,\arabic*},ref={RQ\,\arabic*},itemsep=0ex,topsep=2.25pt}

\begin{document}

\title{\large\textbf {WIP: Identifying Tutorial Affordances for Interdisciplinary Learning Environments}}

\author{\IEEEauthorblockN{Hannah Kim}
\IEEEauthorblockA{\textit{Department of Biology}\\
\textit{Temple University}\\
\textit{Philadelphia, PA, USA}\\
hannah.kim0007@temple.edu}
\and
\IEEEauthorblockN{Sergei L. Kosakovsky Pond}
\IEEEauthorblockA{\textit{Department of Biology}\\
\textit{Temple University}\\
\textit{Philadelphia, PA, USA}\\
spond@temple.edu}
\and
\IEEEauthorblockN{Stephen MacNeil}
\IEEEauthorblockA{\textit{Dept. of Computer \& Information Sciences}\\
\textit{Temple University}\\
\textit{Philadelphia, PA, USA}\\
stephen.macneil@temple.edu}
}

\maketitle

\begin{center}
    \rule{8.4cm}{0.1pt}\\[1ex]
    \fontsize{7.5pt}{9.5pt}\selectfont
    \copyright 2024 IEEE. Personal use of this material is permitted. Permission from IEEE must be obtained for all other uses, in any current or future media, including reprinting/republishing this material for advertising or promotional purposes, creating new collective works, for resale or redistribution to servers or lists, or reuse of any copyrighted component of this work in other works.\\ doi:10.1109/FIE61694.2024.10893187\\
    \rule{8.4cm}{0.1pt}\\[1ex]
\end{center}
\begin{abstract}
\small 
This work-in-progress research paper explores the effectiveness of tutorials in interdisciplinary learning environments, specifically focusing on bioinformatics. Tutorials are typically designed for a single audience, but our study aims to uncover how they function in contexts where learners have diverse backgrounds. With the rise of interdisciplinary learning, the importance of learning materials that accommodate diverse learner needs has become evident. We chose bioinformatics as our context because it involves at least two distinct user groups: those with computational backgrounds and those with biological backgrounds. The goal of our research is to better understand current bioinformatics software tutorial designs and assess them in the conceptual framework of interdisciplinarity. We conducted a content analysis of 22 representative bioinformatics software tutorials to identify design patterns and understand their strengths and limitations. We found common codes in the representative tutorials and synthesized them into ten themes. Our assessment shows degrees to which current bioinformatics software tutorials fulfill interdisciplinarity.
\end{abstract}

\IEEEpeerreviewmaketitle

\section{Introduction}
Interdisciplinarity involves solving problems by integrating ideas and approaches from multiple disciplines \cite{klein1990interdisciplinarity, borrego2010definitions}. Although the cost of merging ideas from more than one discipline is steep, it is justified by the significant impact of technological or socio-technological problems being addressed (e.g. climate change and public health) \cite{borrego2010definitions}. A key challenge in interdisciplinary research is the establishment of common ground between disciplines. The challenge entails the necessity of exploring another discipline for concepts and methods when required by the problem. 

Bioinformatics is an interdisciplinary field in which the goal of research is to find important biological patterns using computational and statistical methods. Bioinformatics software users can range from those in experimental biology \cite{giardine2005galaxy, jalili2020galaxy} to those in software development \cite{mulder2018development}. Bioinformatics-oriented courses are offered across many disciplines, including biology, biotechnology, life sciences, pharmacy, computer science, and information technology \cite{magana2014survey}. Naturally, users come from vastly different backgrounds. However, the time allotted for bioinformatics coursework in a formal training is often limited \cite{mulder2018development}. Hence, there is a constant need for users to learn outside formal learning environments. 

Informal learning can supplement formal learning \cite{carliner2013have}. After completing formal education, individuals continue to gain knowledge through informal learning for the rest of their lives. Afforded by the prevalence of e-learning, researchers routinely encounter tutorials online. In bioinformatics, many tutorials focus on providing information about bioinformatics software.

Developing tutorials in bioinformatics presents unique challenges due to its interdisciplinary nature. Most software and corresponding tutorials are built on the assumption that users have basic scripting or data-cleaning skills. In reality, users vary widely, ranging from those without much computational experience to those capable of developing their own software. Hence, there must be an indication of design deficiencies for specific user groups.

To understand the affordances of bioinformatics software tutorials, we used the conceptual framework of interdisciplinarity proposed by Borrego and Newswander \cite{borrego2010definitions}. Their five criteria used to assess interdisciplinary work are: (1) grounding in traditional disciplines; (2) synergistic integration of disciplinary insights; (3) communication and translation (C\&T) across boundaries to reconcile disciplinary differences; (4) critical awareness of achievable scopes and limitations; and (5) teamwork among researchers.

Previously, users of bioinformatics tool documentation were categorized as new, experienced, power, and all \cite{karimzadeh2018top}. Bioinformatics software documentation for new users encompasses the manuscript, readme, quickstart, reference manual, FAQ, and searchable forum or mailing list, with the first three exclusively reserved for new users. However, this approach has limitations. First, the taxonomy above ignores the diverse user disciplinary backgrounds in bioinformatics \cite{borrego2010definitions}. Second, the static nature of academic publications does not reflect the dynamic nature of software development expected in successful tools \cite{gardner2022sustained}. Third, academic publications are not always open access and therefore not always accessible to users. Hence, we defined a bioinformatics software tutorial as a standalone set of documentation containing information about the biological and methodological significance and design that supports new users. With this definition of a tutorial, we asked the following research questions:

\begin{RQ} 
    \item What features do bioinformatics software tutorials have?
    \item What do the tutorial designs achieve in the interdisciplinary learning environment?
\end{RQ}

\section{Content Analysis Methodology} 

We adopted a content analysis approach using bioinformatics software tutorials to systematically explore design patterns. 

\subsection{Sampling Methods}
We obtained a list of 22 representative bioinformatics software tools from a 2018 study \cite{karimzadeh2018top}. We identified their tutorial websites by querying the software names with the term ``bioinformatics'' on Google. We gathered tutorial materials for each software from the corresponding website. The samples were collected between May 1, 2024, and May 12, 2024.

 \noindent \textbf{Software List}: BLAST \cite{altschul1990basic}, MEGA \cite{kumar1994mega}, PLINK \cite{purcell2007plink}, SWiss-PdbViewer \cite{guex1996swiss}, SAMtools \cite{li2011statistical}, BWA \cite{li2009fast}, EMBOSS \cite{rice2000emboss}, Bowtie 2 \cite{langmead2012fast}, DESeq 2 \cite{love2014moderated}, Cufflinks \cite{trapnell2010transcript}, GATK \cite{mckenna2010genome}, limma \cite{ritchie2015limma}, edgeR \cite{robinson2010edger}, MACS \cite{zhang2008model}, Bedtools \cite{quinlan2010bedtools}, Clustal Omega \cite{sievers2011fast}, Meme Suite \cite{bailey2009meme}, Trimmomatic \cite{bolger2014trimmomatic}, STAR \cite{dobin2013star}, Segway \cite{hoffman:unsupervised}, Bioconductor \cite{gentleman2004bioconductor}, and Picard Tools \cite{Picard2019toolkit}.

\subsection{Coder} 
The main coder was a bioinformatics researcher with a background in chemistry, computational biology, and higher education teaching. Given the coder’s hybrid academic background and a decade of experience in the field, they were considered well-qualified to extract themes comprehensively.

\subsection{Open Coding Process}
The coder used open coding \cite{williams2019art} inductively to capture emerging themes in tutorials. Open coding was accompanied by memoing to extract meanings effectively \cite{birks2008memoing}. After identifying open codes from all tutorials, we iterated through the tutorials again deductively to record the presence of each code in each tutorial. We organized the records in the form of tables. The table was filled primarily using Boolean values (e.g., command line tool availability), but also including numeric (e.g., year of first publication) and categorical (e.g., Python or R package manager) values. We also collected codes describing new user materials (e.g., from Software A, ``Help; Getting Started; Manual; Guide; Handbook''). Throughout this process, the codes were discussed with the research team and adjusted as needed through a mediation process. 

\subsection{Analyzing Codes}
\subsubsection{Quantitative analysis}
We converted each code row into binary or numeric values as appropriate. Based on this data, we examined the frequency, mean, and standard deviation of each open code. We also investigated the correlations between the codes. Preliminary statistical analyses were performed using the R programming language.

\subsubsection{Qualitative analysis}
We implemented an axial coding approach to identify relationships among open codes \cite{williams2019art}. We focused on outcomes achieved by each code. 

\subsubsection{Evaluation}
We calculated the arithmetic mean of code frequencies to assess the interdisciplinarity of bioinformatics software tutorials. Codes were assigned to each interdisciplinarity criterion with redundancy: if a code supported the criterion, it was given a (+) value; if it contradicted the criterion, it was given a (-) value.

\section{Preliminary Results}
\subsection{Open Coding}
\subsubsection{At a Glance}
From the open coding of tutorial materials of the bioinformatics software (N=22), we obtained 56 open codes. We verified that the software list was representative, covering bioinformatics topics extensively, from phylogenetics to transcriptomics to repository management. The year of first software availability ranged from 1990 to 2015, with a median of 2010. These tools were each maintained for 4 to 33 years (mean=14.2, sd=6.8), based on their first availability and last release dates. All tutorials had the following codes: general software significance explained, download/installation information found easily, command-line tools, current version information and instructions written by the developers. 

\subsubsection{Tutorials Come in Various Names}
Unlike what the study in 2018 \cite{karimzadeh2018top} suggested, the vocabulary around new user materials was not straightforward. In fact, materials for new users were frequently found in different documents of the same software. There were between 2 and 16 different words per software signaling new user materials (mode=6.0, sd=3.7). Out of 48 words describing new user materials, 18 of them occurred more than twice across different software. \\
\textbf{Top 18 words}: (manual,12), (documentation,11), (FAQ,11), (help,11), (tutorial,9), (gettingstarted,8), (guide,8), (quickstart,8), (workflow,7), (reference,5), (resource,4), (vignette,4), (description,3), (glossary,3), (how,3), (note,3), (tips\&tricks,3), and (troubleshooting,3).

\subsubsection{Correlations}
The amount of tutorial material was expected to increase with the length of maintenance. We observed that the years of software maintenance and the count of codes describing new user materials showed a moderate correlation (Pearson's r=0.45). When we limited the same analysis to tools that have basic tutorials across more than one platform, we found the correlation to be negligible (r=0.06; N=10). On the other hand, when we limited the analysis to the tools that have basic tutorials on only one platform, there was a strong correlation (r=0.63; N=12). Based on our memo, most tutorials using more than one platform utilized platforms with infrastructure such as GitHub \cite{Github}, Readthedocs \cite{Readthedocs}, SourceForge \cite{SourceForge}, and Bioconductor \cite{gentleman2004bioconductor}. Each platform had a set of default functionalities that can aid in tutorial development. We counted any personal website as one platform. In summary, the volume of new user materials increased over time when tutorials were maintained on a single platform. 

We investigated the presence of discussion group promotions in tutorials over time. We found a moderate correlation between developer-promoted discussion groups and the year of first availability (r=0.52). Seventeen out of nineteen tools that were first available after the year 2000 had a discussion group link in their tutorials. Other time-related codes, such as the years of software maintenance (r=-0.37) and the year of latest software release (r=0.29), did not show a strong correlation with the discussion group code.

\subsection{Axial Coding}
By grouping open codes, we aimed to understand recurring themes. Ten axial codes have been synthesized from the 51 open codes. Each axial code had 3 to 8 open codes involved. From this point forward, codes will be referenced by their numerical identifiers as detailed in Table \ref{table:1}. Note that some open codes in X1, X2, and X4 are likely reserved for a specific group of new users.

\begin{table}[]
    \centering
    \begin{tabular}{l|l|l}
       \hline 
          \textbf{Codes}  & \textbf{\%} & \% Visualization 
   \\
    \hline 
    \hline 
   \textbf{X1. User Interface}  & \textbf{} & 
   \\
   \hline 
   C1. Web application & 41\% & \newmoon\newmoon\newmoon\newmoon\fullmoon\fullmoon\fullmoon\fullmoon\fullmoon\fullmoon
   \\
   C2. Graphical user interface on local machine & 27\% & \newmoon\newmoon\newmoon\fullmoon\fullmoon\fullmoon\fullmoon\fullmoon\fullmoon\fullmoon
   \\
   C3. Command-line tool & 100\% & \newmoon\newmoon\newmoon\newmoon\newmoon\newmoon\newmoon\newmoon\newmoon\newmoon
   \\
   \hline
   \textbf{X2. Installation}  & \textbf{} & 
   \\
   \hline 
    C4. Download/install information found easily & 100\% & \newmoon\newmoon\newmoon\newmoon\newmoon\newmoon\newmoon\newmoon\newmoon\newmoon
   \\
   C5. Additional installation instruction & 91\% & \newmoon\newmoon\newmoon\newmoon\newmoon\newmoon\newmoon\newmoon\newmoon\fullmoon
   \\
    C6. Operating system requirements & 86\% & \newmoon\newmoon\newmoon\newmoon\newmoon\newmoon\newmoon\newmoon\newmoon\fullmoon
   \\
   C7. Package managers, esp. Python- or R-based & 41\% & \newmoon\newmoon\newmoon\newmoon\fullmoon\fullmoon\fullmoon\fullmoon\fullmoon\fullmoon
   \\
   C8. Pre-compiled binaries & 86\% & \newmoon\newmoon\newmoon\newmoon\newmoon\newmoon\newmoon\newmoon\newmoon\fullmoon
   \\
   C9. Docker image available & 27\% & \newmoon\newmoon\newmoon\fullmoon\fullmoon\fullmoon\fullmoon\fullmoon\fullmoon\fullmoon
   \\
   \hline 
   \textbf{X3. Data Specification}  & \textbf{} & 
   \\
   \hline
   C10. Sample data & 82\% & \newmoon\newmoon\newmoon\newmoon\newmoon\newmoon\newmoon\newmoon\fullmoon\fullmoon
   \\
   C11. Raw sample data with reference & 55\% & \newmoon\newmoon\newmoon\newmoon\newmoon\newmoon\fullmoon\fullmoon\fullmoon\fullmoon
   \\
   C12. Input data format & 91\% & \newmoon\newmoon\newmoon\newmoon\newmoon\newmoon\newmoon\newmoon\newmoon\fullmoon
   \\
   C13. Output description & 68\% & \newmoon\newmoon\newmoon\newmoon\newmoon\newmoon\newmoon\fullmoon\fullmoon\fullmoon
   \\
   \hline
   \textbf{X4. Format}  & \textbf{} &
   \\
   \hline
   C14. Link to video tutorials & 18\% & \newmoon\newmoon\fullmoon\fullmoon\fullmoon\fullmoon\fullmoon\fullmoon\fullmoon\fullmoon
   \\
   C15. Clear list of tasks \& options in tutorial & 95\% & \newmoon\newmoon\newmoon\newmoon\newmoon\newmoon\newmoon\newmoon\newmoon\newmoon
   \\
   C16. Tutorial curated by topic & 82\% & \newmoon\newmoon\newmoon\newmoon\newmoon\newmoon\newmoon\newmoon\fullmoon\fullmoon
   \\
   C17. Step-by-step instruction & 73\% & \newmoon\newmoon\newmoon\newmoon\newmoon\newmoon\newmoon\fullmoon\fullmoon\fullmoon
   \\
   C18. Github or source code & 86\% & \newmoon\newmoon\newmoon\newmoon\newmoon\newmoon\newmoon\newmoon\newmoon\fullmoon
   \\
   C19. Software release notes & 95\% & \newmoon\newmoon\newmoon\newmoon\newmoon\newmoon\newmoon\newmoon\newmoon\newmoon
   \\
   \hline
   \textbf{X5. Relevance}  & \textbf{} &
   \\
   \hline
   C20. Software name explained & 77\% & \newmoon\newmoon\newmoon\newmoon\newmoon\newmoon\newmoon\newmoon\fullmoon\fullmoon
   \\
   C21. General software significance explained & 100\% & \newmoon\newmoon\newmoon\newmoon\newmoon\newmoon\newmoon\newmoon\newmoon\newmoon
   \\
   C22. General software method explained & 82\% & \newmoon\newmoon\newmoon\newmoon\newmoon\newmoon\newmoon\newmoon\fullmoon\fullmoon
   \\
   \hline
   \textbf{X6. Maintenance}  & \textbf{} & 
   \\
   \hline 
   C23. Instructions written by the developers & 100\% & \newmoon\newmoon\newmoon\newmoon\newmoon\newmoon\newmoon\newmoon\newmoon\newmoon
   \\
   C24. Current version information & 100\% & \newmoon\newmoon\newmoon\newmoon\newmoon\newmoon\newmoon\newmoon\newmoon\newmoon
   \\
   C25. Years of software maintenance (mean=14.18) & NA & NA
   \\
   C26. Citation information conspicuous & 91\% & \newmoon\newmoon\newmoon\newmoon\newmoon\newmoon\newmoon\newmoon\newmoon\fullmoon
   \\
   C27. Citation info less accessible e.g., in FAQ & 18\% & \newmoon\newmoon\fullmoon\fullmoon\fullmoon\fullmoon\fullmoon\fullmoon\fullmoon\fullmoon
   \\
   C28. Page link not working & 41\% & \newmoon\newmoon\newmoon\newmoon\fullmoon\fullmoon\fullmoon\fullmoon\fullmoon\fullmoon
   \\ 
   \hline 
   \textbf{X7. Resilience}  & \textbf{} & 
   \\
   \hline 
   C29. Last tutorial update date & 59\% & \newmoon\newmoon\newmoon\newmoon\newmoon\newmoon\fullmoon\fullmoon\fullmoon\fullmoon
   \\
   C30. Software limitations & 59\% & \newmoon\newmoon\newmoon\newmoon\newmoon\newmoon\fullmoon\fullmoon\fullmoon\fullmoon
   \\
   C31. Statement of error or retraction & 5\% & \newmoon\fullmoon\fullmoon\fullmoon\fullmoon\fullmoon\fullmoon\fullmoon\fullmoon\fullmoon
   \\
   C32. Deprecation information & 41\% & \newmoon\newmoon\newmoon\newmoon\fullmoon\fullmoon\fullmoon\fullmoon\fullmoon\fullmoon
   \\
   C33. Latest software release (mean=2021.59) & NA & NA
   \\
   C34. Multiple software versions on the main page & 9\% & \newmoon\fullmoon\fullmoon\fullmoon\fullmoon\fullmoon\fullmoon\fullmoon\fullmoon\fullmoon
   \\
   C35. Tutorials for different major software versions & 27\% & \newmoon\newmoon\newmoon\fullmoon\fullmoon\fullmoon\fullmoon\fullmoon\fullmoon\fullmoon
   \\
   C36. Guide to landing pages itself & 9\% & \newmoon\fullmoon\fullmoon\fullmoon\fullmoon\fullmoon\fullmoon\fullmoon\fullmoon\fullmoon
   \\
   \hline 
   \textbf{X8. Promotion of Community}  & \textbf{} & 
   \\
   \hline
   C37. Instructions written by community members & 32\% & \newmoon\newmoon\newmoon\fullmoon\fullmoon\fullmoon\fullmoon\fullmoon\fullmoon\fullmoon
   \\
   C38. User rating & 14\% & \newmoon\fullmoon\fullmoon\fullmoon\fullmoon\fullmoon\fullmoon\fullmoon\fullmoon\fullmoon
   \\
   C39. User feedback form & 77\% & \newmoon\newmoon\newmoon\newmoon\newmoon\newmoon\newmoon\newmoon\fullmoon\fullmoon
   \\
   C40. Use cases via featuring user work & 41\% & \newmoon\newmoon\newmoon\newmoon\fullmoon\fullmoon\fullmoon\fullmoon\fullmoon\fullmoon
   \\
   C41. Developer-organized social events & 14\% & \newmoon\fullmoon\fullmoon\fullmoon\fullmoon\fullmoon\fullmoon\fullmoon\fullmoon\fullmoon
   \\
   C42. Developer-promoted discussion group & 77\% & \newmoon\newmoon\newmoon\newmoon\newmoon\newmoon\newmoon\newmoon\fullmoon\fullmoon
   \\
   \hline
   \textbf{X9. Willingness to Collaborate}  & \textbf{} &
   \\
   \hline 
   C43. Software dependency on other tools & 36\% & \newmoon\newmoon\newmoon\newmoon\fullmoon\fullmoon\fullmoon\fullmoon\fullmoon\fullmoon
   \\
   C44. Link to references other than its own & 91\% & \newmoon\newmoon\newmoon\newmoon\newmoon\newmoon\newmoon\newmoon\newmoon\fullmoon
   \\
   C45. Instructions for other developers & 36\% & \newmoon\newmoon\newmoon\newmoon\fullmoon\fullmoon\fullmoon\fullmoon\fullmoon\fullmoon
   \\
   C46. Instructions for administrators & 9\% & \newmoon\fullmoon\fullmoon\fullmoon\fullmoon\fullmoon\fullmoon\fullmoon\fullmoon\fullmoon
   \\
   C47. Copyright Information & 100\% & \newmoon\newmoon\newmoon\newmoon\newmoon\newmoon\newmoon\newmoon\newmoon\newmoon
   \\
   \hline 
   \textbf{X10. Marketing}  & \textbf{} & 
   \\
   \hline 
   C48. Ad/news & 41\% & \newmoon\newmoon\newmoon\newmoon\fullmoon\fullmoon\fullmoon\fullmoon\fullmoon\fullmoon
   \\
   C49. Logo & 41\% & \newmoon\newmoon\newmoon\newmoon\fullmoon\fullmoon\fullmoon\fullmoon\fullmoon\fullmoon
   \\
   C50. Performance comparison with competitors & 18\% & \newmoon\newmoon\fullmoon\fullmoon\fullmoon\fullmoon\fullmoon\fullmoon\fullmoon\fullmoon
   \\
   C51. Summary statistics of download & 32\% & \newmoon\newmoon\newmoon\fullmoon\fullmoon\fullmoon\fullmoon\fullmoon\fullmoon\fullmoon
    \\
    \hline
    \end{tabular}
    \vspace{1pt}
    \caption{Bioinformatics Software Tutorial Open Codes by Themes}
    \label{table:1}
\end{table}

\subsection{Evaluation of Interdisciplinarity}
We assigned codes to the criteria previously proposed \cite{borrego2010definitions}.
\begin{itemize}[label={}]
  \item \verb|+| Disciplinary grounding: \{C3, C7-C9, C18-C19\}
  \item \verb|-| Disciplinary grounding: $\emptyset$
  \item \verb|+| Integration: $\emptyset$
  \item \verb|-| Integration: $\emptyset$
  \item \verb|+| C\&T: \{C1-C2, C4-C6, C10-C13, C14-C17,
  \\C20-C22, C23-C24, C26, C37-C42, C48-C51\}
  \item \verb|-| C\&T: \{C27-C28\}
  \item \verb|+| Critical Awareness: \{C29-C32\}
  \item \verb|-| Critical Awareness: \{C34-36\}
  \item \verb|+| Teamwork: \{C43-47\}
  \item \verb|-| Teamwork: $\emptyset$
\end{itemize}

We assessed each criterion by taking the arithmetic mean.
\begin{itemize}
  \item Disciplinary grounding: \verb|73%|
  \item Integration: \verb|NA|
  \item C\&T: \verb|58%|
  \item Critical Awareness: \verb|17%|
  \item Teamwork: \verb|54%|
\end{itemize}

\section{Interpretation}

\subsection{RQ1 $\rightarrow$ Features in Bioinformatics Software Tutorials}
We captured ten themes reflecting trends in tutorial features (See Table \ref{table:1}). Tutorials were developed for various types of user interfaces and typically included features related to installation, data specification, format, and relevance. We observed that features related to tutorial maintenance varied, with some indicating effectiveness and others indicating ineffectiveness. Similar mixed signals were noted in features related to resilience. Furthermore, features aimed at promoting community engagement, collaboration, and marketing were commonly found across tutorials.

\subsection{RQ2 $\rightarrow$ Design Achievements by Themes}
We examined the achievements of the tutorial features based on the identified themes. The availability of X1 determined the accessibility of the software and their tutorials. X2-X5 included features dedicated to providing structures. Users with no prior knowledge of a bioinformatics tool can benefit from having structures \cite{tennyson1991structured}. For example, providing clear installation and data specification information can directly influence user engagement with the software. Descriptions of general software significance (C21) and method (C22) can motivate new users to further explore the software. Incorporating multimedia presentations in the form of videos (C14) can also improve new user learning experience \cite{tani2022can}. X6-X10 were themes that required time to establish. Specifically, X6-X7 captured how the software and their tutorials accommodated changes or imperfections. Good maintenance of tutorials can serve as an indicator of software integrity. Resilience showed how tutorials have managed the accumulation of knowledge over time. X8-X10 were environmentally conscious themes. These features demonstrated an understanding of the user community and collaborators. The effort to distinguish the software in the market was a natural outcome of this awareness.

Not all features have uniformly achieved usefulness for all users. There were three different types of user interfaces: two graphical user interfaces (C1-C2) and one command-line interface (C3). The availability of graphical user interfaces varied dramatically across tutorials (41\% and 27\%), while the command-line interface was universally available. Even the union of graphical user interface features was only 55\%. Developing command-line tools is less expensive in many ways. Nevertheless, using command-line tools requires additional steps for users compared to using tools with graphical user interfaces. This apparent trend in tools, which have been maintained for an average of 14.18 years (C25), reflects unrealistic developer expectations about the majority of potential users. The learning cost associated with these expectations is likely prohibitive to many users in the interdisciplinary environment.

Having a large volume of tutorial content may not necessarily improve learning outcomes. Because we sampled tools by representativeness, their tutorials showed a significant correlation between the duration of maintenance (C25, mean=14.18) and the number of words used to describe new user materials (not shown in the table, mode=6). While a rich collection of learning materials can benefit users from different disciplines, using six different words to indicate new user materials is excessive \cite{dicarlo2009too}, especially in e-learning environments where personalized guidance is often lacking. This highlights the need for greater tutorial resilience to changes (X7), as having too many pointers can undermine efforts to provide structures.

Promotion of community engagement can address knowledge gaps. Many recent bioinformatics tutorials have directed users towards a community (r=0.52, 17 out of 19 after 2010). Developer promotion of community (X8) aligns with a constructivist approach \cite{bada2015constructivism}. While a structured approach is effective for beginners \cite{tennyson1991structured}, it becomes less effective when developer expectations of users inadvertently diverge from reality. By promoting knowledge sharing and increasing interaction opportunities in a community, developers can accommodate diverse user needs in interdisciplinary learning environments.

Some bioinformatics software tutorials draw advantages from leveraging default infrastructures. For example, Bioconductor \cite{gentleman2004bioconductor} requires developers to maintain tutorial features such as ``vignettes" and keep track of the date of the last tutorial update (C29). Conda \cite{conda, cock2009biopython} not only simplifies software dependency management but also makes installation (C7) straightforward. Additionally, package management platforms often require features such as the latest software version (C24) and release (C33). Overall, default infrastructures significantly facilitate the management of new user materials. 

\subsection{RQ2 $\rightarrow$ Design Achievements by Interdisciplinarity}

We assessed the interdisciplinarity of bioinformatics software tutorials based on the five criteria proposed by Borrego and Newswander. We found a strong indication of disciplinary grounding (73\%). Communication and translation across disciplinary boundaries, as well as teamwork, were moderately implemented (58\% and 54\%). Critical awareness was rarely observed (17\%). To understand the synergy of integrating multiple disciplines, it is necessary to evaluate outcomes achieved beyond traditional disciplinary boundaries; however, this aspect was outside the scope of our content analysis.

Interdisciplinarity evaluation supplemented the thematic analysis. Each theme could be better understood through the lens of interdisciplinarity. As observed across X1, X2, and X4, many features were grounded in computational disciplines. Information related to other disciplines (e.g., biology) was present throughout the tutorials, but no features could be specifically assigned to these disciplines. Features related to communication and translation were observed across almost all themes, although a couple of them were counterproductive. This trend indicates the widespread, though moderate, presence of such efforts. Features related to critical awareness were identical to those of resilience but could be further categorized into favorable and unfavorable sets. All features related to willingness to collaborate were interpreted as teamwork-related.

\section{Conclusion}
We have identified design features in bioinformatics software tutorials and assessed their achievements. Integrating ideas and approaches from multiple disciplines is challenging yet highly rewarding. By analyzing current affordances, we can detect strengths and weaknesses to leverage in interdisciplinary learning environments. In bioinformatics, there is currently a noticeable disciplinary skew in tutorials intended for learners from diverse backgrounds. Recognizing such design deficiencies will improve the development of learning materials in environments beyond traditional learning settings.

\section{Potential Limitations}
Although we have systematically studied tutorial affordances, some limitations may affect our interpretation. First, only a single coder coded the data. Second, the sample size was small (N=22). Third, our representative samples came from an older study in 2018. Because bioinformatics is a fast-moving field, the tutorial landscape may have shifted since then. Fourth, our findings assume the popularity of the tools. What we found may not apply to less popular tools. Nevertheless, our work-in-progress study provides sufficient insights into the affordances of tutorials in interdisciplinary learning environments for future works.

\section*{Acknowledgment}
The authors would like to thank H. Ji and Dr. M. Yeon for providing feedback during the conception of the work. The authors would also like to thank C. Zastudil, S. Bernstein, K. Angelikas, and C. Kapp for stimulating discussions around the topic during the weekly HCI Reading Group meetings.


\end{document}